\documentclass[aps,prl,twocolumn,showpacs,superscriptaddress,reprint]{revtex4}
 \usepackage{amsmath}
\usepackage[usenames,dvipsnames]{color}
\usepackage[normalem]{ulem}
 \usepackage{ulem}

\newcommand{\kb}{k_{\mbox{\scriptsize B}}}
\newcommand{\dd}{{\mbox{d}}}
\newcommand{\lgle}{\left\langle}
\newcommand{\rgle}{\right\rangle}
\newcommand{\diff}[2]{\dfrac{\mbox{d} #1}{\mbox{d} {#2} } }
\newcommand{\pdif}[2]{\frac{\partial #1}{\partial #2 } }
\newcommand{\rev}{{(\mbox{\scriptsize rev})}}
\newcommand{\irr}{{(\mbox{\scriptsize irr})}}

\newcommand{\ud}{{(\mbox{\scriptsize ud})}}
\newcommand{\e}{\mbox{\large e}}
\newcommand{\Rot}{{\mbox{\scriptsize R}}}
\newcommand{\sol}{{\mbox{\scriptsize solvent}}}
\newcommand{\col}{{\mbox{\scriptsize colloid}}}

 \newcommand{\new}[1]{{#1}}
 \newcommand{\delete}[1]{}
 \newcommand{\renew}[1]{{#1}}
 \newcommand{\newnew}[1]{{#1}}

\begin{document}

\title{The Entropy Anomaly and the Linear Irreversible Thermodynamics}

\author{Kunimasa Miyazaki}
\affiliation{Department of Physics, Nagoya University, Nagoya, 464-8602, Japan}
\email[Corresponding author: ]{miyazaki@r.phys.nagoya-u.ac.jp}
\author{Yohei Nakayama}
\affiliation{Department of Physics, Faculty of Science and Engineering, Chuo University, Kasuga, Tokyo 112-8551, Japan}
\author{Hiromichi Matsuyama}
\affiliation{Department of Physics, Nagoya University, Nagoya, 464-8602, Japan}

\date{\today}

\begin{abstract}
The irreversible currents and entropy production rate of a dilute colloidal suspension
are calculated using \delete{the} linear irreversible
thermodynamics and the linear response theory. 
The ``anomalous'' or ``hidden'' entropy \delete{recently discussed}\new{that has been the subject of
 recent discussion} in the context of \delete{the} stochastic
 thermodynamics is fully accounted \new{for} in these \delete{classic}\new{classical} frameworks.
We show that the two distinct formulations lead to identical results as long as 
the local equilibrium assumption\new{,} or equivalently the linear response theory, is valid.
\end{abstract}

 \maketitle

 \delete{Last}\new{In the last} several decades, we have witnessed \new{an} enormous development of thermodynamics 
of systems far from equilibrium and with small sizes~(see
Refs.~\cite{Seifert2012,Sekimoto2010book,Klages2012book} and references therein).
The concept\new{s} of \delete{the} irreversibility and entropy \delete{has}\new{have} been
completely \delete{renewed}\new{reformulated} in the context of \delete{the} 
stochastic thermodynamics and \new{the} fluctuation theorems, in which  
the trajectories of fluctuating observables are directly linked to \delete{the} entropy
production\delete{s} and the \delete{extent}\new{quantification} of \delete{the} irreversibility.
Brownian motion of a colloidal particle has often been employed as the simplest archetypal model to
formulate these theories and \new{to} demonstrate their consequences~\cite{Klages2012book}.
Usually, the overdamped Langevin equation\new{,} \delete{in which}\new{which neglects} the effect of
\delete{the} inertia\renew{,} \delete{is neglected} is used 
since the time scale of \delete{the} relaxation of the momentum is negligibly small in most realistic situations.
However, \delete{it has been demonstrated recently}\new{recent studies have demonstrated} 
that the overdamped \renew{Langevin equation} cannot correctly describe the
entropy production rate even in the simplest nonequilibrium system of a single Brownian particle in
a constant temperature
gradient~\cite{Celani2012prl,Bo2014jsp,Spinney2012pre,Kawaguchi2013pre,Ge2014pre,Ge2015jsmte,Lan2015sr}.
These studies have shown that the true entropy production rate can be obtained by incorporating the
momentum as well as the position of the particle into the stochastic equation. In other words, the
underdamped Langevin equation has to be used to describe the entropy production even in the limit of
small mass or inertia.
As the entropy production cannot be captured by only using the slow variable\delete{s}\new{,} or the
position of the particle \delete{which}\new{that} is observed in the usual experimental setting, it
has been dubbed the ``anomalous''~\cite{Celani2012prl} or ``hidden'' entropy~\cite{Kawaguchi2013pre}. 
This anomalous entropy production \new{rate} is written as~\cite{Celani2012prl}
\begin{equation}
\diff{\lgle S_{\mbox{\scriptsize anom}} \rgle}{t} = \frac{(d+2)}{6}\int\dd\vec{r}~\rho
 \frac{\kb^2}{\zeta T}|\nabla T|^2,
\label{eq:Celani1}
\end{equation}
where $\rho$ is the probability density of the particle position, $d$ is the spatial dimension\delete{s},
$\zeta$ is the friction coefficient of the particle, $\kb$ is the Boltzmann constant, and \(T\) is an inhomogeneous temperature field.
More recently, the formulation is generalized to a particle with \delete{the} rotational degree\new{s} of freedom
and \delete{under}\new{in a} convection~\cite{Lan2015sr}.

On the other hand, the entropy production of macroscopic nonequilibrium systems is often described by
\delete{the} classic\new{al} linear irreversible thermodynamics (LIT). The theory is based on the \delete{prerequisite} 
assumptions that  
the system is \delete{at} locally \new{at} equilibrium and \new{that} the irreversible currents are linearly proportional 
to thermodynamic forces~\cite{degrootmazur1962}. 
The relationship between \delete{the} mesoscopic stochastic thermodynamics and macroscopic LIT remains somewhat elusive.
Although Eq.~(\ref{eq:Celani1}) is obviously \delete{the} quadratic \delete{form of} \new{in} thermodynamic forces 
as prescribed by LIT~\cite{Stolovitzky1998pla,Ge2015jsmte}, 
the equivalence of the two descriptions and its validity are yet to be clarified.

In this \delete{short} paper, we show that the anomalous entropy production\delete{s}
given by Eq.~(\ref{eq:Celani1}) and the results of Refs.~\cite{Ge2015jsmte,Lan2015sr}
are accounted \new{for} \delete{only} using \new{only} \delete{the} LIT and \delete{the} 
linear response theory, or the Green-Kubo formula,  
both of which are established as the hydrodynamic description of the nonequilibrium systems. 
The result demonstrates the equivalence between the stochastic and linear irreversible thermodynamic
descriptions as long as the local equilibrium \delete{and linear response} condition\delete{s} \delete{are}\new{is}
satisfied\delete{and we focus on the average value of the entropy productions}. 
Macroscopic LIT \delete{description} suffices to describe the \new{average of the} entropy production up to the quadratic order in
$\nabla T$.
 \new{The benefit and advantage of stochastic thermodynamics over LIT are the capability
 of going beyond the local equilibrium regime, where the effects of inertia and 
nonlinearity of thermodynamic forces become relevant~\cite{Spinney2012pre}.  
In addition, the stochastic thermodynamics can describe the fluctuations
 of the entropy directly in terms of the particle trajectories, which is impossible in LIT. }
The starting point of \delete{the} LIT is the local thermodynamic relation
\begin{equation}
 \dd s = \sum_{\alpha}\pdif{s}{\varphi_\alpha}\dd \varphi_{\alpha},
\label{eq:S1}
\end{equation}
where $s= s(\vec{r}, t)$ is the entropy per unit volume of the system, 
$\varphi_\alpha(\vec{r}, t)$ \delete{are}\new{is} the local extensive variable\delete{s,} and 
$\partial s/\partial \varphi_\alpha$ \delete{are}\new{is} the
conjugate intensive variable\delete{s}. 
\delete{Essential is that Eq.~(\ref{eq:S1}) is assumed to hold}
\new{An essential assumption is that Eq.~(\ref{eq:S1}) holds}
 locally so that the system at each position
relaxes \delete{fast enough}\new{sufficiently quickly} to an equilibrium state defined by the local variables. 
Usually, the slow and conserved quantities such as the density, total momentum, and total energy
fields have to be employed as $\varphi_\alpha$~\cite{degrootmazur1962}. 
Since $\varphi_\alpha$ \delete{are} \new{is a} conserved variable\delete{s}, its time evolution is written
using the current $\vec{J}_{\alpha}$ as
$\partial{\varphi_\alpha}/\partial t = -\nabla\cdot\vec{J}_{\alpha}$.
Furthermore, $\vec{J}_{\alpha}$ can be decomposed into the reversible and irreversible parts as 
$\vec{J}_{\alpha}= \vec{J}_{\alpha}^{~\rev}+\vec{J}_{\alpha}^{~\irr}$ and the irreversible current is
written as 
\begin{equation}
\vec{J}_{\alpha}^{~\irr} = \sum_{\beta}L_{\alpha\beta}\vec{X}_\beta,  
\label{eq:irr1}
\end{equation}
where $\vec{X}_\beta = \nabla (\partial s/\partial \varphi_{\beta})$ is the thermodynamic force and
$L_{\alpha\beta}$ is the Onsager coefficient~\cite{onsager1931a}.
Plugging \delete{the above} \new{these} expressions \delete{to}\new{into} Eq.~(\ref{eq:S1})\new{,} we obtain the entropy production \new{rate}
of the whole system $S = \int\dd\vec{r}~s$ expressed as
\begin{equation}
\diff{S}{t} = \int\dd\vec{r}~\sum_{\alpha\beta}\vec{X}_{\alpha}L_{\alpha\beta}\vec{X}_{\beta} \geq 0,
\label{eq:EPR}
\end{equation}
which is the second law of thermodynamics.
On the other hand, the linear response theory, or the Green-Kubo formula, claims that the Onsager
coefficient is microscopically expressed in terms of the two point correlation functions at
equilibrium as 
\begin{equation}
L_{\alpha\beta} = \frac{1}{\kb V}\int_{0}^{\infty}\dd t~  \lgle J_{0\alpha}^{\irr}(t)J_{0\beta}^{\irr}(0)\rgle,
\label{eq:Onasger}
\end{equation}
where $V$ is the volume of the system and $J_{0\alpha}^{\irr}(t)$ is the microscopic irreversible
current fluctuation\delete{s} integrated over the volume of the system~\cite{kubo1978,hansen1986}. The
Onsager reciprocal relation $L_{\alpha\beta}=L_{\beta\alpha}$ is a direct 
consequence of the microscopic time reversibility of the correlation function in 
Eq.~(\ref{eq:Onasger}).

Let us apply \delete{the theory}\new{LIT} \delete{for} \new{to} a dilute \new{suspension of
noninteracting} Brownian particle\new{s} \delete{systems,}  
\delete{suspended} \new{diffusing} in a fluid in the presence of \delete{the} \new{a} temperature gradient. 
\delete{We consider an ideal colloidal suspension where the interparticle
interactions are neglected. }
We regard the system as \delete{the}\new{a} two\renew{-}component mixture \delete{consisting} \new{which consists}
of \delete{the  colloidal} \new{Brownian} particles and \delete{the surrounding} solvent
\delete{fluid} \new{molecules}. 
The relevant thermodynamic variables are the number density of \delete{the} particles and the 
\new{total} energy density $\varphi= (\rho, e)$.
 Note that the density of the solvent \delete{fluid} \new{molecules} does not need to be considered as the total momentum
 flow is conserved~\cite{degrootmazur1962}.  
We \delete{address} \new{emphasize} that $e$ is the {\it total} energy density of both the
\new{Brownian} particles and the solvent\delete{fluid}. The energy of each component is not the
conserved quantity as they can exchange their \delete{momentum}\new{momenta} through the friction. 
The time evolution of $\rho$ and $e$ \delete{are} \new{is} written as
$\partial \rho/\partial t= -\nabla\cdot\vec{J}_{\rho}$ and
$\partial e/\partial t= -\nabla\cdot\vec{J}_{e}$. 
The total energy current, $\vec{J}_e$, contains the contributions both of the \delete{colloidal} \new{Brownian} particles
and the solvent\delete{fluid}. 
Here, we assume 
the contribution \delete{from} \new{of} the solvent\delete{s} to be well decoupled from
\delete{those} \new{that} of the \delete{colloids} \new{Brownian particles}, so that 
we may write $\vec{J}_e= \vec{J}_{e,\sol}+ \vec{J}_{e,\col}$.
Hereafter, we shall only consider the contribution of \delete{colloids} \new{Brownian particles} and simply denote\delete{s}
$\vec{J}_{e,\col}$ as $\vec{J}_{e}$.
In the absence of \delete{the macroscopic} convection, both $\vec{J}_\rho$ and $\vec{J}_{e}$ are purely
irreversible currents and, according to Eq.~(\ref{eq:irr1}), \delete{they} are \delete{written
as}\new{given by}
\begin{equation}
 \left\{
\begin{aligned}
&
 \vec{J}_\rho = L_{\rho\rho}\nabla \left( -\frac{\mu}{T}\right) + L_{\rho e}\nabla \left(\frac{1}{T}\right),
\\
&
 \vec{J}_e = L_{e\rho}\nabla \left( -\frac{\mu}{T}\right) + L_{ee}\nabla \left(\frac{1}{T}\right),
\end{aligned}
       \right.
\label{eq:JJ1}
\end{equation}
where $\mu$ is the chemical potential.
$L_{\rho e}$ and $L_{e \rho}$ are the cross coefficients whose ratios with $L_{\rho\rho}$ are often called the Soret and Dufour
coefficients~\cite{degrootmazur1962}. 
In the case of the \new{ideally} dilute \delete{ideal colloidal} suspension, one can infer $L_{\rho e}$ (and thus
$L_{e\rho}$ from the reciprocal relation) from thermodynamic argument without 
resorting to the Green-Kubo formula. 
The chemical potential for the dilute \delete{system} \new{suspension (an ideal\new{, classical} gas of colloids)} is  
\begin{equation}
 \mu = \kb T \left( \ln \rho  - \frac{d}{2}\ln T + C\right),
\label{eq:CP1}
\end{equation}
where $C$ is a constant~\new{\cite{Hill1987book}. }
If the system is confined in a box and is in the stationary state, $\vec{J}_{\rho}=0$ and 
the (osmotic) pressure must be spatially uniform, {\it i.e.}, $\nabla (\rho \kb T)=0$.
Comparing this \delete{equation with}\new{to} Eq.~(\ref{eq:JJ1}) \delete{together with}\new{and using} 
the chemical potential of Eq.~(\ref{eq:CP1}), we arrive at 
\begin{equation}
\frac{L_{\rho e}}{L_{\rho\rho}}= \frac{(d+2)\kb T}{2} \equiv h, 
\label{eq:GD1}
\end{equation}
where $h$\delete{$=(d+2)\kb T/2$} is the partial enthalpy.
\delete{We address}\new{Note} that the Soret coefficient for \delete{the}\new{this} system is \delete{purely} a
\new{purely} thermodynamic quantity.
Using this result, $\vec{J}_{\rho}$ is written as 
\begin{equation}
 \vec{J}_{\rho} = -D \nabla \rho - \frac{\rho D}{T}\nabla T.
\label{eq:GD2}
\end{equation}
Here $D= \kb  L_{\rho\rho}/\rho$ is the diffusion coefficient.
In the dilute limit, $\rho$ is proportional to the probability distribution of a single
Brownian particle and $D$ becomes the self-diffusion coefficient which is given by
the Einstein relation, $\kb T/\zeta$\delete{,}\new{.} \delete{and}\new{Thus,} Eq.~(\ref{eq:GD2})
\delete{becomes}\new{is} equivalent \delete{with}\new{to} the
Smoluchowski equation for a single particle under a 
temperature gradient~\cite{vanKampen1988ibm,widder1989,Sancho1982}.

\delete{Now we}\new{We} demonstrate that \delete{the} all \delete{matrix} elements of the Onsager matrix
can be derived from the Green-Kubo formula given by Eq.~(\ref{eq:Onasger}).
The microscopic irreversible currents of the density and energy are given by~\cite{vanKampen1991jsta}
\begin{equation}
\begin{aligned}
&
\vec{J}_{0\rho}^{\irr}(t) = \sum_{k=1}^{N} \vec{v}_k (t),
\hspace*{0.5cm}
\vec{J}_{0 e}^{\irr}(t) = \sum_{k=1}^{N} \frac{m}{2}\delete{|v_k^2(t)|}\renew{|v_k(t)|^2}\vec{v}_k (t),
\end{aligned}
\label{eq:Jq2}
\end{equation}
where \new{$N$ is the number of particles, $m$ is the mass, and} $\vec{v}_{k}(t)$ \delete{and $m$
are} \new{is} the velocity \delete{and mass} of the $k$-th particle\delete{, respectively}.
Using the velocity correlation function of \delete{colloidal} \new{Brownian} particles given by 
\begin{equation}
 \lgle v_{k i}(t)v_{l j} (0)\rgle = \frac{\kb
  T}{m}\delta_{kl}\delta_{ij}\e^{-\gamma t},
\end{equation}
where $\gamma \equiv \zeta/m$ and the 
letters \(i\) and \(j\) represent the coordinate, the Green-Kubo formula readily reproduces the aforementioned result\delete{;}\new{:}
\begin{equation}
 L_{\rho\rho}=  \frac{\rho T}{\zeta} = \frac{\rho D}{\kb}. 
\end{equation}
\renew{The} other coefficients can also be calculated by invoking the fact that $\vec{v}_k(t)$ 
follows \delete{the} Gaussian statistics and thus all higher moments can be decomposed into \delete{the} multiples of
the second moment\delete{s}. For example,
$\lgle v_{kx}(t) v_{kx}^3(0)\rgle = 3\lgle v_{kx}(t)v_{kx}(0)\rgle \lgle v_{kx}^2(0)\rgle$.
After \new{a} straightforward calculation\delete{s}, we  obtain $L_{\rho e}=L_{e\rho}= (d+2)\rho\kb T^2/2\zeta = h
L_{\rho\rho}$ which reconfirms the result of Eq.~(\ref{eq:GD1}) and thus Eq.~(\ref{eq:GD2}). 
Note that Eq. (\ref{eq:GD2}) was originally derived  by adiabatic elimination of 
the momentum from the Kramers equation~\cite{vanKampen1988ibm,widder1989} or by taking the
overdamped limit of the underdamped 
Langevin equation\newnew{~\cite{footnote1},} 
taking special care with the interpretation of the
multiplicative noise~\cite{Sancho1982}.
It is interesting that the same equation can be derived from the Green-Kubo formula.

Likewise, $L_{ee}$ \delete{is}\new{can be} \delete{given by} \new{calculated as}
\begin{equation}
\begin{aligned}
 L_{ee} 
&
= 
\frac{1}{\kb V}\int_{0}^{\infty}\dd t~  \lgle J_{0e x}(t)J_{0 e x}(0)\rgle
\\ 
&
= 
\rho \lambda T^2+  h L_{\rho e},
\\ 
\end{aligned}
\label{eq:lambda0}
\end{equation}
where  $\lambda$  is the thermal conductivity given by
\begin{equation}
 \lambda = \frac{(d+2)\kb^2 T}{6\zeta}.
\label{eq:lambda1}
\end{equation}
Using these results, the energy current can be written as the sum of the enthalpy current and pure heat
flow current as 
\begin{equation}
 \vec{J}_{e} = h \vec{J}_\rho  +  \vec{J}_q, 
\label{eq:Je1}
\end{equation}
where $\vec{J}_q = \vec{J}_e  - h \vec{J}_{\rho} = - \rho \lambda \nabla T$.
This \delete{is an} expression \new{is} often introduced in textbooks~\cite{degrootmazur1962,Callen1985}. 
Inserting all Onsager coefficients evaluated above \delete{to}\new{into} Eq.~(\ref{eq:EPR}), the entropy production rate 
becomes 
\begin{equation}
\diff{S}{t} = \int\dd\vec{r}~
\left[
\frac{1}{L_{\rho\rho}}\vec{J}_\rho^{~2} 
+ \frac{\rho\lambda}{T^2}   |\nabla {T}|^2
\right].
\label{eq:EPR2}
\end{equation}
The first term on the right hand side \delete{of this equation}\new{here} correspond\new{s} to 
the entropy production rate 
\delete{obtained by applying} 
\new{derived from} \delete{the} stochastic thermodynamics \delete{to}
\new{with the} overdamped Langevin equation~\cite{Seifert2012}\delete{,}\new{.} \delete{and t}
\new{T}he second term is nothing but Eq.~(\ref{eq:Celani1}), the result \delete{reported}
\new{derived} by Celani {\it et al.} \new{using stochastic thermodynamics with the underdamped Langevin equation~\cite{Celani2012prl}.}
Of course, 
\delete{the entropy production of the whole system of the colloids and solvent 
fluid which play a role of the heat bath for the colloids should be given by the sum 
of Eq.~(\ref{eq:EPR2}) and the contribution from the heat flow of the solvent which is expressed in
the same form as Eq.~(\ref{eq:Celani1}) but with $\rho \lambda$ for the solvent fluid. }
\new{the total entropy production rate of the whole system is given by adding to the right hand side
of Eq.~(\ref{eq:EPR2}) the contribution of the heat flow by the solvent which is expressed in the same form as 
Eq.~(\ref{eq:Celani1}) but with $\rho \lambda$ of the solvent.}
\new{With these results in mind, it is natural that the Brownian particles convey heat 
and thus generate entropy even when there is no mass flow ($\vec{J}_{\rho}=0$), 
in much the same way that the solvent fluid does.  
In this sense, it may be somewhat misleading to call this entropy production rate ``anomalous''~\cite{Celani2012prl}.}

The above argument can be readily extended to \delete{the case in the presence of}
\new{Brownian particles with} \delete{the} rotational degree\new{s} of
freedom \delete{for colloidal particles}~\cite{Lan2015sr}. 
For \delete{simplification} \new{simplicity}, we assume that the \delete{colloidal} \new{Brownian} particle is spherical and 
therefore the rotational \delete{dynamics is}\new{degrees of freedom are} not coupled to the
translational one\new{s}~\cite{Lan2015sr,Dhont1996}. 
As the number of degree\new{s} of freedom\delete{s are} \new{is} doubled, the partial enthalpy is replaced by 
$h = (d + 1)\kb T$ and $d/2\ln T$ in the chemical potential, Eq.~(\ref{eq:CP1}), is replaced by $d\ln T$. 
Most importantly, the irreversible energy current \delete{should} now include\new{s}
\delete{the}\new{an} extra energy term due to \delete{the} 
rotation and is given by 
\begin{equation}
\vec{J}_{0 e}\renew{(t)} 
= \sum_{k=1}^{N} \left(\frac{m}{2}\delete{|v_k^2|}\renew{|v_k(t)|^2}
+ \frac{I}{2}\delete{|\omega_k^2|}\renew{|\omega_k(t)|^2}\right)\vec{v}_k \renew{(t)}
\end{equation}
instead of Eq.~(\ref{eq:Jq2}), where $I$ is the \new{moment of} inertia \delete{moment} and $\vec{\omega}_k$ is the angular
\delete{momentum}\new{velocity}.  
The Langevin equation for the angular momentum is \delete{basically} the same as \delete{that for}\new{in} the translational
\delete{one}\new{case}, except that $m$ and $\zeta$ are replaced by $I$ and the rotational friction coefficient
$\zeta_\Rot$, respectively. Therefore, the \delete{time} \new{$\vec{\omega}_k$} correlation function
\delete{of $\vec{\omega}_k$} is written as  
\begin{equation}
 \lgle \omega_{k x}(t)\omega_{k x} (0)\rgle = \frac{\kb T}{I}\e^{-\gamma_\Rot t}, 
\end{equation}
where $\gamma_\Rot = \zeta_{\Rot}/I$.
Calculation of the Onsager coefficients by the Green-Kubo formula is straightforward. 
$L_{\rho\rho}$ and $L_{\rho e}=L_{e\rho}$ are unchanged. 
For $L_{ee}$, due to the extra contribution from the rotational energy, $\lambda$ is replaced by $\lambda+\lambda_{\Rot}$ with
\begin{equation}
 \lambda_{\Rot} = \frac{d\kb^2 T}{m}\frac{1}{\gamma + 2\gamma_\Rot}.
\label{eq:lambdaRot}
\end{equation}
This is the result reported in Ref.~\cite{Lan2015sr}.

Another interesting case is \delete{the entropy production due to the viscous flow}
\new{the system in the presence of convective flow}~\cite{Lan2015sr}. 
The entropy production \new{rate of incompressible viscous flow} \delete{for 
an incompressible system} 
is written as~\cite{degrootmazur1962,landau1959} 
\begin{equation}
\diff{S}{t} = \int\dd\vec{r}~\frac{\eta}{2T}\left( \pdif{u_i}{x_j}+\pdif{u_j}{x_i}\right)^2,
\label{eq:EPR4}
\end{equation}
where $\eta$ is the shear viscosity and $u_i$ is the velocity field.
It should be emphasized that both $\eta$ and $u_i$ are the quantities of the whole system\new{,
which comprises} of the mixture of 
\delete{the colloids} \new{Brownian particles} and solvent.
Again\new{,} if the thermodynamics and dynamics of \delete{the solvent and colloids} \new{the Brownian
particles and solvent} are well decoupled, we can 
write $\eta = \eta_{\sol} + \eta_{\col}$ and hereafter 
we shall consider only the latter, which shall be denoted as $\eta$.
The Green-Kubo formula for the shear viscosity is~\cite{hansen1986}
\begin{equation}
 \eta =  \frac{1}{\kb TV}\int_{0}^{\infty}\dd t~  \lgle \sigma_{0 xy}(t)\sigma_{0xy}(0)\rgle,
\end{equation}
where $\sigma_{0xy}\renew{(t)}$ is the stress tensor fluctuation given by
$\sigma_{0xy}\renew{(t)}= \sum_{k}mv_{kx}\renew{(t)}v_{ky}\renew{(t)}$. 
Unless the rotational degrees of freedom couple to the translational one, the stress correlation function is easily
calculated to give 
\begin{equation}
 \eta =  \frac{m\rho\kb T}{2\zeta}.
\end{equation}
Eq.~(\ref{eq:EPR4}) together with this expression is the result reported in Ref.~\cite{Lan2015sr}. 

\delete{These}\new{The above} results \delete{shown above} claim that the \new{average of the}
entropy production rate evaluated from  \delete{the trajectories of
a colloidal particle} \new{the underdamped stochastic
thermodynamics~\cite{Celani2012prl,Bo2014jsp,Spinney2012pre,Kawaguchi2013pre,Ge2014pre,Ge2015jsmte,Lan2015sr}}
 \delete{in the full phase space} is equivalent \delete{from} \new{to} that obtained from \delete{the} LIT\delete{as an average}.
\delete{We emphasize that the LIT is based on the overdamped
dynamics of the slow conserved variables}
\new{We emphasize that LIT is valid at slow time scales where the overdamped approximation is valid} 
as \delete{it} is obvious from the diffusion equation given by Eq.~(\ref{eq:GD2})\new{.} \delete{and does not
require}\new{It is not necessary} to explore \delete{the faster} \new{short} time scales in order to retrieve the correct entropy
production \new{rate}.
\new{The reason why LIT explains the entropy production rate correctly at 
the overdamped limit while the overdamped stochastic dynamics does not is because information about the momenta of particles is
correctly taken into account in LIT. For example, the energy density $e$ 
contains the kinetic contribution. 
Moreover, the subtlety that comes into play when bridging the gap between overdamped and underdamped
stochastic thermodynamics in nonequilibrium systems does not appear in  
LIT because the Onsager coefficients such as $D$ and $\lambda$ are calculated using 
only fluctuations {\it at equilibrium}, and the Green-Kubo integrals are not affected by the
time scales of the dynamics~\cite{vanKampen1988ibm,widder1989}.}
The prerequisite for \new{the underdamped} stochastic thermodynamics  to be
equivalent \delete{with} \new{to} \delete{the} LIT is the local equilibrium assumption under which the linear response theory
is validated.
Indeed, one can show that \delete{the} extending the description to \delete{the} shorter time scales does not alter
the entropy production rate as \delete{far}\new{long} as the local equilibrium condition 
\delete{is guaranteed}\new{holds}.
\delete{In the}\new{At} shorter time scales, or in the underdamped \delete{limit}\new{case}, the relaxation of $\vec{J}_{\rho}$ becomes
relevant and, therefore,
$\vec{J}_{\rho}$ should be upgraded \delete{as the relevant} \new{to a} thermodynamic variable\delete{s}. Concomitantly, the total energy should be
redefined as $e = e'+ mJ_\rho^2/2\rho$, as discussed in \delete{CH}\new{Chapter} III. \S 4 of
Ref.~\cite{degrootmazur1962}.
It is natural to expect that the underdamped equation for $\vec{J}_{\rho}$ is obtained by simply
adding the 
inertial term to the overdamped equation, Eq.~(\ref{eq:GD2})~\cite{degrootmazur1962,doi1992};
\begin{equation}
 \frac{m}{\zeta}\pdif{\vec{J}_{\rho}}{t} = -{\vec{J}_{\rho}} +D \nabla \rho + \frac{\rho D}{T}\nabla
  T. 
\label{eq:GD3}
\end{equation}
The irreversible part of \new{the} energy current is now given by
$\vec{J}_q$ and not by $\vec{J}_e$ because the partial enthalpy current, $h \vec{J}_\rho$, in
Eq.~(\ref{eq:Je1}) should be regarded as the reversible current.
Corresponding to this re-definition, the microscopic energy current in Eq.~(\ref{eq:Jq2}) should be
replaced by
$\vec{J}_{0 e}^{\irr}(t) = \sum_{k=1}^{N} (\frac{m}{2}\delete{|v_k^2(t)|}\renew{|v_k(t)|^2}- h)\vec{v}_k (t)$, which leads to
the Onsager coefficient $L_{ee} = \rho \lambda T^2$ instead of Eq.~(\ref{eq:lambda0}).
Therefore, the entropy production rate in \delete{this}\new{the} underdamped dynamics is \delete{written as}
\begin{equation}
\diff{S_{\ud}}{t} = \int\dd\vec{r}~
\left[
\frac{\zeta}{\rho T}\vec{J}_\rho^{~{2}} 
+ 
\frac{\rho \lambda}{T^2} |\nabla T|^2
\right],
\label{eq:EPR5}
\end{equation}
which is identical to the overdamped \delete{counterpart}\new{case}, Eq.~(\ref{eq:EPR2}). 
This result bolsters our conclusion that the overdamped description correctly describe \delete{the}
``anomalous'' entropy production. 
Although appealing, there is \delete{an}\new{a} caveat \delete{which}\new{that} should be addressed.
Eq\new{.}\delete{uation}~(\ref{eq:GD3}) is valid only if the local equilibrium condition is satisfied. 
But in most realistic situations, 
the relaxation time $\gamma^{-1}= m/\zeta$ is comparable with or shorter than the time
scale\delete{s} \new{required} for \delete{the} local equilibration
to be achieved.

The importance of the local equilibrium assumption upon which \delete{the} LIT is based cannot be overstated.
As \delete{it is}\new{was} clearly demonstrated by Widder \delete{{\it et al.}}\new{and
Titulaer}~\cite{widder1989} for the same system \delete{we} 
\new{as the one} considered \new{in this paper}, the local equilibrium condition is violated if the parameter
 \begin{equation}
\varepsilon \equiv    \frac{1}{\gamma}\sqrt{\frac{\kb T}{m}}\times \frac{|\nabla T|}{T}
 \end{equation}
is \delete{finite}\new{large}.  
In \delete{the limit of}\new{the lowest order in} $\varepsilon$\delete{$\rightarrow 0$}, 
their result leads to the same conclusions \delete{of}\new{as} this
paper and Ref.~\cite{Celani2012prl}. 
The \new{higher order} corrections due to finite $\varepsilon$ correspond to the higher order terms
in the $\gamma^{-1}$ expansion  
of the Kramers equation~\cite{widder1989,Stolovitzky1998pla}, or the Burnett corrections. 
The correction\new{s} \delete{has}\new{have} been evaluated and assessed by Spinney {\it et al.}~\cite{Spinney2012pre}. 
The situations where $\varepsilon$ is \delete{finite}\new{large} would be exactly the realm where the stochastic
thermodynamic description with the full-phase description of the momentum and position is
indispensable and beneficial.  
But otherwise, the linear irreversible thermodynamics with the linear response theory (the
Green-Kubo formula) would suffice as demonstrated in this paper. 

Lastly, \delete{the}\new{an} external conserved force can be also included in the formulation discussed in this paper. 
One \delete{may}\new{can} simply add a potential 
$\phi(x)$ to the enthalpy and the chemical potential. \delete{But}\renew{However,} the spatial variance of $\phi(x)$ should
be small enough not to undermine the local equilibrium condition.

\begin{acknowledgments}
We thank Shin-ichi Sasa\new{,} \delete{and} Masato Itami\new{, and John Wojdylo} for fruitful
 discussion \new{and valuable comments}.
KM acknowledge KAKENHI No. 
16H04034, 
25103005, 
25000002. 
YN acknowledge Grant-in-Aid for Young Scientists (B) No.
17K14373.
HM acknowledges MEXT, Program for Leading Graduate Schools ``IGER''.
\end{acknowledgments}


\begin{thebibliography}{23}
\bibitem{footnote1}
\newnew{In the overdamped limit, only the first
order contribution in $1/\gamma$, or more rigorously in $\varepsilon= \gamma^{-1}\sqrt{\kb T/m}
|\nabla T|/T$, is retained in the $\gamma^{-1}$-expansion of the Kramers equation or in the
adiabatic elimination of the momentum in the underdamped Langevin equation.}
\end{thebibliography}
\end{document}